\documentclass[12pt]{article}
\usepackage{amsmath,amsfonts}
\usepackage{bm}

\def\d{\partial}
\def\bra{\langle}
\def\ket{\rangle}
\newcommand{\R}{{\bm R}}
\def\eqref#1{(\ref{#1})}

\def\cH{{\cal H}}
\def\cD{{\cal D}}
\def\cN{{\cal N}}
\def\dk#1#2{\frac{ d^{#2}{#1} }{ (2\pi)^{#2} }} 
\def\da#1#2{\frac{ d{#1}}{{#1}^{{#2}+1}}}
\def\bvec#1{{\bm #1}}
\def\vb{\bvec{b}}
\def\vk{\bvec{k}}
\def\vq{\bvec{q}}
\def\vx{\bvec{x}}
\begin{document}
\title{Wavelet based regularization for Euclidean field theory\thanks{Proc.
 GROUP 24: Physical and Mathematical Aspects of 
Symmetries, edited by J-P. Gazeau, R. Kerner, J-P. Antoine, S.
Metens, J-Y. Thibon; IOP Publishing, Bristol, 2003}}
\author{M V Altaisky \\ 
Joint Institute for Nuclear Research, Dubna, 141980, Russia;
\\ and Space Research Institute, Moscow, 117997, Russia;\\ 
{\em e-mail}: altaisky@mx.iki.rssi.ru}
\date{Sep 20, 2002}
\maketitle
\begin{abstract}
It is shown that Euclidean field theory with polynomial interaction, can be regularized using the wavelet representation of the fields.
The connections between wavelet based regularization and stochastic quantization are considered.
\end{abstract}

\section{Introduction}
The connections between quantum field theory and stochastic
differential equations have been calling constant attention for
quite a long time \cite{Nelson,GJ1981}.
We know that
stochastic processes often posses self-similarity.
The renormalization procedure used in quantum field theory is also based
on the self-similarity.
So, it is natural to use for the regularization of field
theories the wavelet transform (WT), the decomposition
with respect to the representation of the affine group.
In this paper two ways of regularization are presented.
First, the direct substitution
of WT of the fields into the action functional leads to
a field theory with scale-dependent coupling constants. Second, the
WT, being substituted into the Parisi-Wu stochastic quantization
scheme  \cite{PW1981},
provides a stochastic regularization with no  extra vertexes introduced into the theory.

\section{Scalar field theory on affine group}
The  Euclidean field theory is defined on $\R^d$ by the generating
functional
\begin{equation}
W_E[J] = \cN \int \cD\phi \exp\left[-S[\phi(x)] + \int  d^dx J(x)\phi(x) \right]
\label{gf},
\end{equation}
where $S[\phi]$ is the Euclidean action.
In the simplest case of a scalar field with the fourth power interaction
\begin{equation}
S[\phi] =\int d^d x
\frac{1}{2}(\d_\mu\phi)^2+ \frac{m^2}{2}\phi^2 + \frac{\lambda}{4!}\phi^4.
\label{f4l}
\end{equation}
The $\phi^4$ theory is often referred to as the Ginsburg-Landau model
for its
ferromagnetic applications. The $\phi^3$ theory is also a useful model.

The perturbation expansion generated by the functional \eqref{gf}
is usually evaluated in $k$-space.
The reformulation of the theory from the coordinate ($x$) to momentum
($k$) representation is a particular case of decomposition
of a function with respect to the representation of a Lie group $G$.
$G: x'\!=\!x\!+\!b$ for the case of Fourier transform, but other groups
may be used as well.
For a locally compact Lie group $G$ acting transitively
on the Hilbert space $\cH$
it is possible to decompose state vectors with respect to the
representations of $G$
\cite{Carey76,DM1976}
\begin{equation}
|\phi\ket = C_\psi^{-1} \int_G U(g)|\psi\ket d\mu(g)\bra\psi| U(g)|\phi\ket.
\label{pu}
\end{equation}
The constant $C_\psi$ is determined by the norm of the
action of $U(g)$ on the fiducial vector $\psi\in\cH$,
$
C_\psi = \|\psi \|^{-2} \int_{g\in G} |\bra\psi|U(g)|\psi \ket|^2 d\mu(g);
$  $d\mu(g)$ is the left-invariant measure on $G$.

Using decomposition \eqref{pu},
it is possible to define a field theory on
a non-abelian Lie group.
Let us consider the fourth power interaction model
$$\int V(x_1,x_2,x_3,x_4)
\phi(x_1)\phi(x_2)\phi(x_3)\phi(x_4)dx_1 dx_2 dx_3 dx_4.$$
Using the notation
$U(g)|\psi\ket \equiv |g,\psi\ket, \quad \bra\phi|g,\psi\ket \equiv \phi(g),
\quad \bra g_1,\psi|D|g_2,\psi\ket \equiv D(g_1,g_2),
$
we can rewrite the generating functional \eqref{gf} in the form
\begin{eqnarray}
\nonumber W_G[J] = \int \cD\phi(g) \exp\Bigl(
-\frac{1}{2}\int_G \phi(g_1)D(g_1,g_2)\phi(g_2)d\mu(g_1) d\mu(g_2) \\
\nonumber -\frac{\lambda}{4!}\int_G V(g_1,g_2,g_3,g_4)
\phi(g_1)\phi(g_2)\phi(g_3)\phi(g_4)
d\mu(g_1) d\mu(g_2)d\mu(g_3) d\mu(g_4)  \\
+ \int_G J(g)\phi(g)d\mu(g)
\Bigr),
\label{gfi4}
\end{eqnarray}
where $V(g_1,g_2,g_3,g_4)$ is the result of
the transform $\phi(g) := \int\overline{U(g)\psi(x)} \phi(x)dx$
applied to $V(x_1,x_2,x_3,x_4)$ in all arguments $x_1,x_2,x_3,x_4$.

Let us turn to the particular case of the affine group.
\begin{equation}
x'=a x+b,\quad U(g)\psi(x)=a^{-d/2}\psi((x-b)/a)), \quad x,x',b \in \R^d
\label{ag}.
\end{equation}
The scalar field $\phi(x)$ in the action
$S[\phi]$ can be written in the form of wavelet decomposition
\begin{equation}
\begin{array}{lcl}
\phi(x) &=& C_\psi^{-1}\int \frac{1}{a^{d/2}}\psi\left(\frac{x-b}{a}\right)\phi_a(b)
\frac{dadb}{a^{d+1}},\\
\phi_a(b) &=& \int  \frac{1}{a^{d/2}}\bar\psi\left(\frac{x-b}{a}\right)\phi(x)d^dx.
\end{array}
\label{wt}
\end{equation}
In the scale-momentum $(a,k)$ representation
the matrix element of the  free field inverse propagator
$
\bra a_1,b_1;\psi | D | a_2, b_2; \psi\ket =
\int \dk{k}{d} e^{ik(b_1-b_2)} D(a_1,a_2,k)
$
has the form
\begin{equation}
D(a_1,a_2,k) = a_1^{d/2} \overline {\hat \psi(a_1 k)} (k^2+m^2)
                 a_2^{d/2}\hat \psi(a_2 k).
\end{equation}
The field theory \eqref{gfi4} with the propagator $D^{-1}(a_1,a_2,k)$ gives
standard Feynman diagram technique, but with extra
  wavelet   factor $a^{d/2}\hat \psi(a k)$  on each line and
the integrations
over the measure $d\mu(a,k)= \dk{k}{d}\da{a}{d}$ instead of $\dk{k}{d}$.

Recalling the power law
dependence  of the coupling constants on the cutoff momentum resulting
from the Wilson expansion,
we can define a scalar field
model on the affine group, with the coupling constant
dependent on scale. Say, the fourth power interaction
can be written as
\begin{equation}
V[\phi] =\int \frac{\lambda(a)}{4!} \phi^4_a(b) d\mu(a,b),
\quad  \lambda(a)\sim a^\nu.
\label{vint}
\end{equation}

The one-loop order contribution to the Green function $G_2$
in the theory with interaction \eqref{vint}
can be evaluated \cite{Alt2001}
by integration over $z = ak$:
\begin{equation}
\int \frac{a^\nu a^d |\hat\psi(ak)|^2 }{k^2+m^2}\dk{k}{d}\da{a}{d}
=   \int \dk{k}{d} \frac{C_\psi^{(\nu)}k^{-\nu}}{k^2+m^2},
C_\psi^{(\nu)}=\int |\hat\psi(z)|^2 \frac{dz}{z^{1-\nu}}.
\end{equation}
Therefore, there are no UV divergences for $\nu>d-2$.
This is a kind of asymptotically free
theory which is hardly appropriate, say, to spin systems. What is required
to get a finite theory  is an interaction vanishing outside a given
domain of scales. Such model is presented in the next section by
means of the stochastic quantization framework.

\section{Stochastic quantization with wavelets}
Let us remind the basic ideas of the stochastic quantization
\cite{PW1981,ZJ1986,DH87,Namiki92}. Let $S[\phi]$ be an action of
the field $\phi(x)$. Instead of calculation of the physical Green
functions, it is possible to introduce the  ``extra-time''
variable $\tau$: $\phi(x) \to \phi(x,\tau)$ and evaluate the
moments $\langle \phi(x_1,\tau_1) \ldots \phi(x_m,\tau_m)
\rangle_\eta$ by averaging over the random process
$\phi(x,\tau,\cdot)$ governed by the Langevin equation with the
Gaussian random force
\begin{equation}
\dot\phi(x,\tau) +\frac{\sigma^2}{2}
\frac{\delta S}{\delta\phi(x,\tau)} = \eta(x,\tau),
\langle \eta(x,\tau)\eta(x',\tau') \rangle = \sigma^2
\delta(x-x')\delta(\tau-\tau').
\label{le}
\end{equation}
The physical Green functions are then obtained by taking the
limit $\tau_1=\ldots=\tau_m=T\to\infty$.

The stochastic quantization procedure has been considered as a
perspective candidate for the regularization of gauge theories, 
for it respects local gauge symmetries in a natural way. However 
a $\delta$-correlated Gaussian random noise in the Langevin equation still 
yields sharp singularities in the perturbation theory. For this 
reason a number of modifications based
on the noise regularization
 $\eta(x,\tau) \to
\int dy R_{xy}(\d^2)\eta(y,\tau)$ have been proposed
\cite{BGZ84,BHST87,IP88}.

In this paper, following \cite{Alt1999}, we start with the random processes defined
directly in wavelet space.
The use of the wavelet coefficients instead of
the original stochastic processes provides
an extra analytical flexibility of the
method:  there exist more than one set of random functions
$W(a,\vb,\cdot)$ the images of which have
coinciding correlation functions.
It is easy to check that the random process generated by wavelet
coefficients with the correlation function
$
\bra \widehat W(a_1,\vk_1) \widehat W(a_2,\vk_2)\ket
= C_\psi (2\pi)^{d} \delta^{d}(\vk_1+\vk_2)
a_1^{d+1} \delta(a_1-a_2) D_0
$
has the same correlation function as the white noise has \cite{e5-35}.

As an example, let us consider the Kardar-Parisi-Zhang
equation \cite{KPZ1986}:
\begin{equation}
\dot Z -\nu \Delta Z = \frac{\lambda}{2} (\nabla Z)^2 + \eta .
\label{kpz}
\end{equation}
Substitution of wavelet transform
$$
Z(x)  = C_\psi^{-1} \int  \exp(\imath(\vk\vx-k_0 t)) a^{\frac{d}{2}}
\hat\psi(a\vk) \hat Z(a,k) \dk{k}{d+1} \da{a}{d}
$$
into \eqref{kpz},
with the random force of the form
\begin{equation}
\bra \widehat \eta(a_1,k_1) \widehat \eta(a_2,k_2)\ket
=  C_\psi (2\pi)^{d+1} \delta^{d+1}(k_1+k_2)
a_1^{d+1} \delta(a_1-a_2)  D(a_2,k_2),
\label{sfnc}
\end{equation}
leads to the integral equation
$$
\begin{array}{lcl}
(-\imath\omega + \nu\vk^2) \hat Z(a,k) &=& \eta(a,k) - \frac{\lambda}{2}
a^{\frac{d}{2}}  \overline{\hat\psi(a\vk)}
C_\psi^{-2} \int (a_1 a_2)^\frac{d}{2}\hat\psi(a_1\vk_1)
\hat\psi(a_2(\vk-\vk_1)) \\ & & \vk_1  (\vk-\vk_1)
\hat Z(a_1,k_1) \hat Z(a_2,k-k_1) \dk{k_1}{d+1} \da{a_1}{d}\da{a_2}{d}.
\end{array}
$$
From this the one-loop contribution the stochastic Green function
follows:
\begin{equation}
\begin{array}{lcl}
G(k) &=& G_0(k) - \lambda^2 G_0^2(k) \int \dk{k_1}{d+1} \Delta(k_1)\\
& & \vk_1 (\vk-\vk_1)|G_0(k_1)|^2 \vk\vk_1 G_0(k-k_1)
+ O(\lambda^4),
\end{array}
\label{G2}
\end{equation}
where  $G_0^{-1}(k) = -\imath\omega + \nu\vk^2$.
The difference from the standard approach \cite{KPZ1986}
is in the appearance of the effective force correlator
\begin{equation}
\Delta(k) \equiv  C_\psi^{-1} \int \frac{da}{a} |\hat\psi(a\vk)|^2 D(a,\vk),
\label{dak}
\end{equation}
which has the meaning of the effective force averaged over all scales.

Let us consider a single-band forcing \cite{e5-35}
$
D(a,\vk)=\delta(a-a_0) D(\vk)
$
and the ``Mexican hat'' wavelet
$
\hat \psi(k) = (2\pi)^{d/2} (-\imath \vk)^2 \exp(-\vk^2/2)
$.
In the leading order in  small parameter
$x=|\vk|/|\vk_1|\ll1$,  the  contribution to the stochastic
Green function is:
$$
G(k) = G_0(k) + \lambda^2 G_0^2(k) \frac{S_d}{(2\pi)^d}
\frac{a_0^3 k^2}{\nu^2}\frac{d-2}{8d}
\int_0^\infty D(\vq)e^{-a_0^2\vq^2}q^{d+1}dq+O(\lambda^4).
$$

\section{Langevin equation for the $\phi^3$ theory with scale-dependent
noise}
Let us apply the same scale-dependent noise \eqref{sfnc} to the Langevin
equation for $\phi^3$ theory.
The standard procedure of the stochastic quantization then comes
from the Langevin equation \cite{IP88}
\begin{equation}
\dot \phi(x,\tau) + \left[
-\Delta \phi + m^2\phi + \frac{\lambda}{2!} \phi^2 \right] = \eta(x,\tau).
\end{equation}
Applying the wavelet transform to this equation, we get:
\begin{equation}
\begin{array}{l}
(-\imath\omega + \vk^2 + m^2)\hat\phi(a,k) =  \hat\eta(a,k)
-\frac{\lambda}{2} a^{\frac{d}{2}}  \overline{\hat\psi(a\vk)}
C_\psi^{-2} \int (a_1 a_2)^\frac{d}{2}\\
\hat\psi(a_1\vk_1) \hat\psi(a_2(\vk-\vk_1))
\hat \phi(a_1,k_1) \hat \phi(a_2,k-k_1) \dk{k_1}{d+1} \da{a_1}{d}\da{a_2}{d}.
\end{array}
\label{phi3w}
\end{equation}
Iterating the integral equation \eqref{phi3w},
we yield the correction to the stochastic Green
function
\begin{equation}
G(k) = G_0(k) + \lambda^2 G_0^2(k) \int \dk{q}{d+1} \Delta(q)
|G_0(q)|^2 G_0(k-q) + O(\lambda^4).
\label{phi3G1}
\end{equation}

The analytical expressions for the stochastic Green functions, can be obtained in
the $\omega\to0$ limit. As an example, we take the
$D(a,\vq)=\delta(a-a_0)D(\vq)$ for $\phi^3$
theory and the Mexican hat wavelet.
The one loop contribution to the stochastic Green function
$G(k) = G_0(k) + G_0^2 \lambda^2 I_3^2 + O(\lambda^4)$ is
$$
\lim_{\omega\to0}I_3^2 = \int \dk{\vq}{d} \Delta (\vq) \frac{1}{2(\vq^2+m^2)}
\cdot\frac{1}{\vq^2+(\vk-\vq)^2+2m^2} \label{i32}.
$$
The same procedure can be applied in
higher loops.
As it can be seen, for constant or compactly supported $D(\vq)$ all
integrals are finite due to the exponential
factor coming from wavelet $\psi$.
\section*{Acknowlegement}
The author is grateful to Profs. H.H\"uffel and V.B.Priezzhev for useful
discussions.

\end{document}